\documentclass[preprint]{aastex}

\shorttitle{Average properties of
  the bow shock}

\begin{document}
\title{Average observed properties of the Earth's quasi-perpendicular
  and quasi-parallel bow shock} \author{A. Czaykowska, T. M.
  Bauer, R. A. Treumann, and W. Baumjohann}
\affil{Max-Planck-Institut f\"ur extraterrestrische Physik, Garching,
  Germany} \email{acz@mpe.mpg.de, thb@mpe.mpg.de, tre@mpe.mpg.de,
  bj@mpe.mpg.de}

\begin{abstract}We present a statistical analysis of 132 dayside (LT
  0700-1700) bow shock crossings of the AMPTE/IRM spacecraft. We
  perform a superposed epoch analysis of plasma and magnetic field
  parameters as well as of low frequency magnetic power spectra some
  minutes upstream and downstream of the bow shock by dividing the
  events into categories depending on the angle $\theta_{Bn}$ between
  bow shock normal and interplanetary magnetic field and on the
  plasma-$\beta$, i.e., the ratio of plasma to magnetic pressure. On
  average, the proton temperature is nearly isotropic downstream of
  the quasi-parallel bow shock ($\theta_{Bn} < 45^\circ$) and it is
  clearly anisotropic with $T_{p\perp}/T_{p\|} \approx 1.5$ downstream
  of the quasi-perpendicular bow shock ($\theta_{Bn} > 45^\circ$). In
  the foreshock upstream of the quasi-parallel bow shock, the power of
  magnetic fluctuations is roughly 1 order of magnitude larger
  ($\delta B \sim 4$\,nT for frequencies 0.01--0.04\,Hz) than upstream
  of the quasi-perpendicular bow shock.  There is no significant
  difference in the magnetic power spectra upstream and downstream of
  the quasi-parallel bow shock, only at the bow shock itself magnetic
  power is enhanced by a factor of 4.  This enhancement may be due to
  an amplification of convecting upstream waves or due to wave
  generation at the shock interface.  On the contrary, downstream of
  the quasi-perpendicular bow shock the magnetic wave activity is
  considerably higher than upstream. Downstream of the
  quasi-perpendicular low-$\beta$ bow shock we find a dominance of the
  left-hand polarized component at frequencies just below the ion
  cyclotron frequency with amplitudes of about 3\,nT. These waves are
  identified as ion cyclotron waves which grow in a low-$\beta$ regime
  due to the proton temperature anisotropy. We find a strong
  correlation of this anisotropy with the intensity of the left-hand
  polarized component. Downstream of some nearly perpendicular
  ($\theta_{Bn} \approx 90^\circ$) high-$\beta$ crossings mirror waves
  are identified. However, there are also cases where the conditions
  for mirror modes are met downstream of the nearly perpendicular
  shock, but no mirror waves are observed.

\end{abstract}


\section{Introduction} 

Since the beginning of the space age, the Earth's bow shock is of
particular interest because it serves as a unique laboratory for the
study of shock waves in collisionless plasmas. Most of our
understanding of structure, dynamics, and dissipation processes of
such shocks has come from in situ spacecraft measurements crossing the
bow shock.  Early observations of waves and particles upstream of the
bow shock can be found in the special issue of the {\it Journal of
  Geophysical Research}, {\it 86}, pp. 4317--4536 [1981].  A
collection of observational and theoretical work on the bow shock is
contained in {\it Stone \& Tsurutani} [1985] and {\it Tsurutani \&
  Stone} [1985].  Reviews focusing on the dissipation processes
taking place at the Earth's bow shock have been given by, e.g., {\it
  Kennel et al}. [1985] and, more recently, by {\it Omidi} [1995].
Plasma wave observations across the bow shock in the high frequency
range have been reviewed by {\it Gurnett} [1985].  {\it Schwartz et
  al}. [1996] have reviewed results concerning low frequency waves in
the magnetosheath region behind the bow shock.

Many of the previous measurements have demonstrated that a large
variety of nonthermal particles is generated at the bow shock.  While
nonthermal electrons can act as a source for high frequency waves,
nonthermal ions can be responsible for low frequency waves.  Electrons
and ions reflected at the shock stream sunward along the
interplanetary magnetic field, thus forming the electron and ion
foreshock, respectively.

Structure, dynamics, and dissipation processes of the bow shock vary
considerably depending on the angle $\theta_{Bn}$ between the upstream
magnetic field and the shock normal, on the plasma $\beta$, i.e., the
ratio of plasma to magnetic pressure in the upstream region, and on
the Mach numbers $M_A$ or $M_{ms}$, i.e., the ratios of the solar wind
velocity along the shock normal to the upstream Alfv\'en or
magnetosonic speed.  For quasi-perpendicular shocks with $\theta_{Bn}
> 45^\circ$, the main transition from the solar wind to the
magnetosheath is accomplished at a sharp ramp.  In contrast,
quasi-parallel shocks with $\theta_{Bn} < 45^\circ$ consist of
large-amplitude pulsations extending into the foreshock region.  For
larger Mach numbers this pulsating structure continuously re-forms by
virtue of collisions between convecting upstream waves and the shock
[{\it Burgess}, 1989] or due to an instability at the interface
between solar wind and heated downstream plasma [{\it Winske et al}.,
1990].

Two-fluid theories of shocks have indicated the presence of a critical
magnetosonic Mach number, $M^*$, above which ion reflection is
required to provide the necessary dissipation. However, it has been
demonstrated by observations [{\it Greenstadt \& Mellott}, 1987; {\it
  Sckopke et al}., 1990] that ion reflection occurs also below $M^*$
and that the distinction between subcritical ($M_{ms}<M^*$) and
supercritical ($M_{ms}>M^*$) shocks is not sharp.  Whereas the ramp of
quasi-perpendicular shocks at high Mach numbers is preceded by a foot
and followed by an overshoot, these features are less prominent at low
Mach numbers [{\it Mellott \& Livesey}, 1987]. While quasi-parallel
shocks are steady at low Alfv\'en Mach numbers, $M_A\le2.3$, they
become unsteady for higher Mach numbers, where they continuously
re-form [{\it Krauss-Varban \& Omidi}, 1991].

An important role in the dissipation process is played by ions
reflected at the bow shock. At quasi-parallel shocks they can escape
from the shock into the foreshock region and drive ion beam
instabilities. These instabilities may excite large-amplitude waves
observed in the foreshock region, e.g., by {\it Le \& Russell} [1992]
and {\it Blanco-Cano \& Schwartz} [1995].  At quasi-perpendicular
shocks the reflected ions gyrate back to the shock and enter the
downstream region, where their presence leads to a strong
perpendicular temperature anisotropy [{\it Sckopke et al}., 1983].
This anisotropy leads to the generation of ion cyclotron and mirror
waves [e.g., {\it Price et al.}, 1986; {\it Gary et al.}, 1993].
These waves have been observed in the Earth's magnetosheath, e.g., by
{\it Sckopke et al.} [1990] and {\it Anderson et al.} [1993,1994].
Closer to the magnetopause the mechanism of field line draping leads
to the formation of anisotropic ion distributions and the formation of
a plasma depletion layer. Waves in this environment have also been
described by {\it Anderson et al.} [1993,1994]. Large-amplitude mirror
waves have been observed in planetary magnetospheres, e.g., by {\it
  Bavassano-Cattaneo et al.} [1998] in Saturn's magnetosphere, where
the ion temperature anisotropies are due to both shock heating and
field line draping.


In the present study we investigate the average behavior of plasma and
magnetic field parameters including the low frequency magnetic wave
power as measured by AMPTE/IRM during a fairly large number of bow
shock crossings. We show that, as expected, quasi-perpendicular and
quasi-parallel bow shocks behave differently even in their average
properties. This difference has been quantified in our investigation.
Section 2 provides a short description of the available data. It is
followed by Section~3 which compares the properties of
quasi-perpendicular and quasi-parallel shocks. In Section 4 low and
high-$\beta$ bow shock crossings are compared for the
quasi-perpendicular shock, and it is outlined why a classification by
$\beta$ is preferred to a classification by Mach number.  Finally,
Section~5 presents concluding remarks.

\section{Data description} 
The present analysis uses data from the AMPTE/IRM satellite.  From the
periods when the apogee of AMPTE/IRM was on the Earth's dayside
(August -- December 1984 and August 1985 -- January 1986), we have
selected all crossings of the satellite through the Earth's bow shock
in the local time interval 0700 -- 1700, whenever there was a
reasonable amount of data measured on both sides of the bow shock,
i.e., at least 2\,min upstream and 4\,min downstream.  Altogether this
gives 132 events, with some events belonging to multiple crossings due
to the fast movement of the bow shock relative to the slowly moving
satellite.  Due to the satellite's orbital parameters, all crossings
occurred at low latitudes, i.e., in the interval $\pm 30^\circ$ from
the ecliptic plane. We analyze the data from the triaxial fluxgate
magnetometer described by {\it L\"uhr et al.} [1985] which gives the
magnetic field vector at a rate of 32 samples per second. In addition,
we use the plasma moments calculated from the three-dimensional
particle distribution functions measured once every spacecraft
revolution ($\sim 4.3$ s) by the plasma instrument [{\it Paschmann et
  al.}, 1985].

\begin{figure}[h!]
\epsscale{0.5}
\plotone{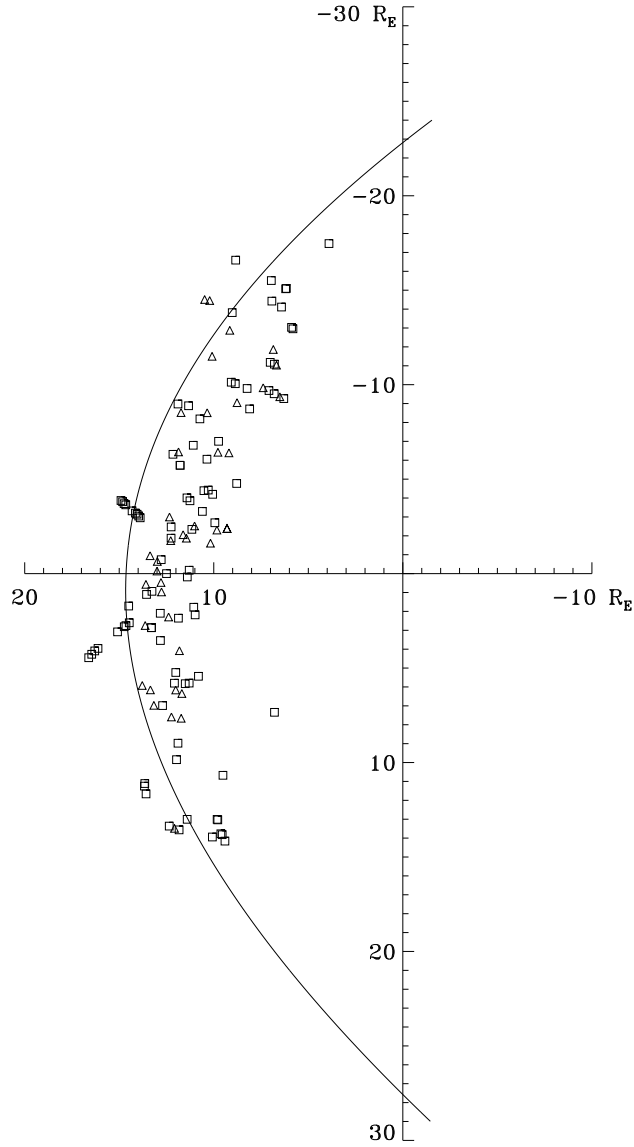}
\caption{\label{positions} GSM-Positions of the 132 AMPTE/IRM bow shock
  crossings rotated to the ecliptic plane
  along meridians. Coordinates are given in Earth radii (R$_E$).
  Quasi-perpendicular crossings are marked with a square,
  quasi-parallel crossings with a triangle. The solid curve represents
  the best fit hyperbola of {\it Fairfield} [1971].}
\end{figure}

In Fig.~\ref{positions} we show the locations of the individual bow
shock crossings rotated into the ecliptic along meridians. Cases where
the angle $\theta_{Bn}$ is less than $45^\circ$, i.e., quasi-parallel
events, and cases where the angle $\theta_{Bn}$ is greater than
$45^\circ$, i.e., quasi-perpendicular events, are shown in addition to
the best fit hyperbola of {\it Fairfield} [1971] using data from the
Imp 1 to 4 and Explorer 33 and 35 spacecraft. It is found that most of
the AMPTE/IRM bow shock crossings occurred closer to the Earth than
Fairfield's average bow shock.  Since we analyze only bow shock
crossings on the dayside, the best fit hyperbola derived from our data
is not reliable at the flanks.  However, the distance of the subsolar
point is well defined. We find a value of 12.3 $R_E$, which is more
than 2 $R_E$ closer to the Earth than the value of 14.6 $R_E$ found by
Fairfield.  In his study, {\it Formisano} [1979] analyzed 1500 bow
shock crossings. He normalized the observed distance $R_{\rm obs}$ of
these crossings to an average value of the solar wind dynamical
pressure using
\begin{equation}
  \label{form}
  R_{\rm norm}=R_{\rm obs}\left( \frac{n_{\rm obs}v_{\rm obs}^2}{n_0v_0^2}\right)^{1/6}
\end{equation}
with a typical value of the solar wind speed $v_0=450$ km/s and
particle density $n_0=9.4$ cm$^{-3}$.  He derived a distance of the
subsolar point of 11.9 $R_E$. Applying the same normalization to the
AMPTE/IRM data, we find a value of 11.7 $R_E$, which is in good
agreement with the result of {\it Formisano} [1979].  This indicates
that the difference of the distance of the subsolar point between
Fairfield's and our study is due to different average solar wind
dynamical pressure.  We interpret this finding as a solar cycle effect
since the AMPTE/IRM data are obtained close to solar activity minimum,
whereas Fairfield's data set is from the years 1964-1968, when solar
activity increased from minimum to maximum.  In solar minimum the
Earth is hit more frequently by high speed solar wind streams than in
solar maximum. The high speed solar wind has, although less dense, a
higher dynamical pressure than the slow solar wind. Hence, the solar
wind dynamical pressure is usually higher on average during solar
minimum than during solar maximum [{\it Fairfield}, 1979]. In
addition, during solar minimum, the heliospheric plasma sheet
described by {\it Winterhalter et al.} [1994] is fairly flat, i.e.,
near the ecliptic plane. With its very high densities it can enhance
the solar wind pressure although the solar wind velocity is only
around 350 km/s.

In a more recent study, {\it Peredo et al.} [1995] investigated 1392
bow shock crossings from 17 spacecraft during the years 1963--1979,
i.e., one and a half solar cycles. They found a dependence on the
Alfv\'en Mach number $M_A$. With the average Alfv\'en Mach number $M_A
= 5.6 \pm 2.9$ of the AMPTE/IRM data set the distance of the subsolar
point should be in the range of 14.0-14.9 $R_E$. Performing a
normalization with the average values of $n_0=7.8$ cm$^{-3}$ and
$v_0=454$ km/s used by {\it Peredo et al.}  [1995], the distance of
the subsolar point of the AMPTE/IRM data set is 12.1 $R_E$. The
results of {\it Peredo et al.} [1995] are thus not in agreement with
our results and those of {\it Formisano} [1979]. {\it Peredo et al.}
[1995] explain this disagreement with the fact that the study of {\it
  Formisano} [1979] is biased by the dominance of the high latitude
HEOS 2 bow shock crossing. However, our data are low-latitude and
agree well with the results of {\it Formisano} [1979].  In principle,
the bow shock position depends on the magnetopause position and on the
standoff distance between the magnetopause and the bow shock. Whereas
the magnetopause position depends only on the solar wind dynamical
pressure, the standoff distance at a given Mach number depends on the
polytropic index $\gamma$ ({\it Spreiter et al.,} 1966). Since our
data were sampled at typical solar minimum conditions, the polytropic
index might be different than in other phases of the solar cycle. This
might contribute to the discrepancy.



\section {Comparison of quasi-perpendicular and quasi-parallel bow
  shock crossings}

We divided the 132 events into 92 quasi-perpendicular ($\theta_{Bn} >
45^\circ$) and 40 quasi-parallel ($\theta_{Bn} <45^\circ$) cases and
compared the average behavior of plasma and magnetic field parameters
and low frequency magnetic fluctuations of the two groups.

Actually, for the quasi-parallel bow shock crossings $\theta_{Bn}$
varies substantially with time in the dynamic foreshock region.  For
these events the angle $\theta_{Bn}$ had to be averaged over a time
interval of about 20 s further upstream to identify them with
quasi-parallel shock crossings. The high level of fluctuations in the
region upstream of the quasi-parallel bow shock is well known [e.g.,
{\it Hoppe et al.}, 1981, {\it Greenstadt et al.,} 1995].

In order to obtain the average behavior of plasma and magnetic field
parameter at the bow shock, one would ideally need average spatial
profiles of these parameters. However, with just one satellite and in
a region with strong plasma flows and strong motions of the region
itself, it is not unambiguously possible to translate the time
profiles into spatial profiles. Therefore we perform a superposed
epoch analysis by averaging time profiles centered on the bow shock
crossing time and consider the result as an approximation for the
average spatial behavior. The time series are aligned on the keytime
with the upstream always preceding, i.e., for outbound crossings the
time sequence had to be reversed.

The keytime, i.e., the bow shock crossing time, is identified with
the steepest drop in the proton velocity. This drop is well defined
for the quasi-perpendicular cases and corresponds, of course, to the
shock ramp.  Due to the large-amplitude pulsations in the foreshock,
the keytime cannot as easily be found in the quasi-parallel cases.
We therefore applied, as an additional criterion for the
quasi-parallel events, that no solar wind-like plasma is allowed to be
visible in the downstream region.  As noted in Section~1,
quasi-parallel shocks consist of large-amplitude pulsations associated
with a sequence of partial transitions from solar wind-like to
magnetosheath-like plasma and vice versa. Thus our definition of the
keytime implies that the keytime of quasi-parallel shocks
corresponds to the downstream end of this pulsating transition region.

We use data from 2\,min upstream to 4\,min downstream for the analysis
of the plasma and magnetic field parameters and data from 3 min
upstream to 9\,min downstream for the low frequency fluctuations,
although not for all events such long time profiles are available.

One has to be aware that superposed epoch analysis can mask small
scale structures, in particular if they are not visible in all events,
like magnetic foot and overshoot structures. These and other features
can be smeared out due to different bow shock velocities with respect
to the satellite for different crossings. This effect becomes worse
the farther away from the keytime the data are averaged. Nevertheless,
superposed epoch analysis is useful to reveal the average plasma and
magnetic field parameters and the typical features in the vicinity of
the key-structure as has been shown by, e.g., {\it Paschmann et al.}
[1993], {\it Phan et al.} [1994], and {\it Bauer et al.} [1997] in
there studies at the magnetopause.

\begin{figure}[h!]
\epsscale{0.75}
\plotone{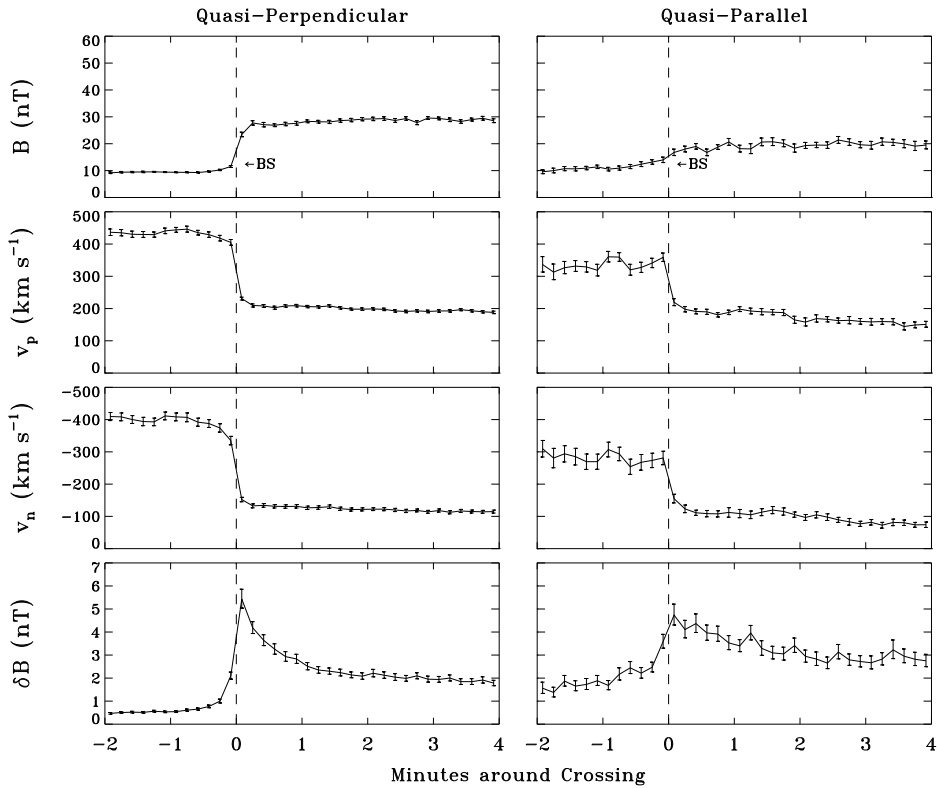}
\caption{Superposed-epoch analysis of plasma and magnetic
  field parameters from 2\,min upstream to 4\,min downstream of the bow
  shock (BS) of 92 quasi-perpendicular (left) and 40 quasi-parallel
  (right) bow shock crossings. Shown are from top to bottom the
  magnitude $B$ of the magnetic field, the magnitude $v_p$ of the proton
  velocity, the proton velocity $v_n$ parallel to the bow shock
  normal vector, and the root mean square amplitude $\delta B_2$ of
  magnetic fluctuations.\label{ppsp1}}
\end{figure}

\begin{figure}[h!]
\epsscale{0.75}
\plotone{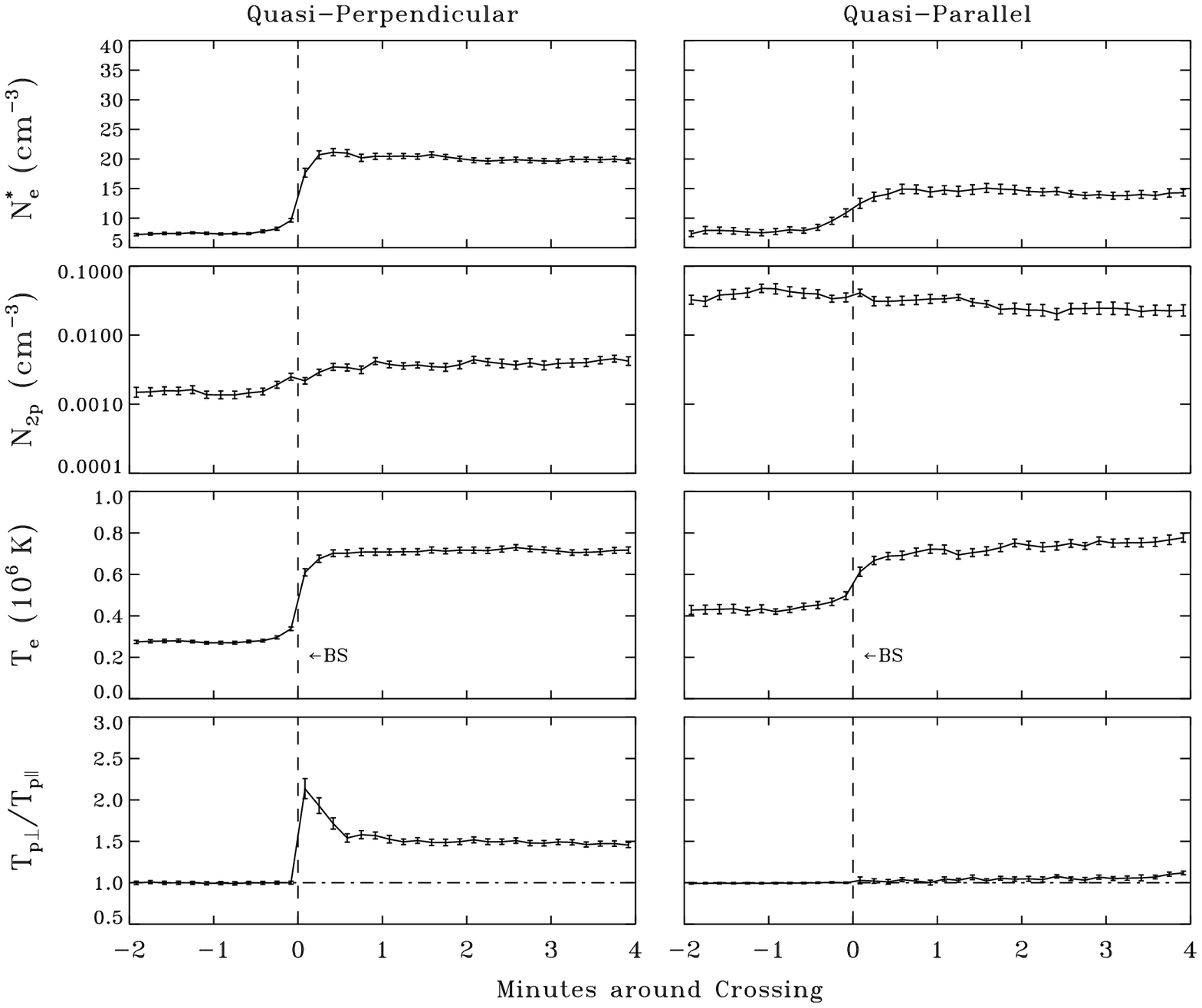}
\caption{\label{ppsp2} Superposed-epoch analysis of plasma
  parameters from 2\,min upstream to 4\,min downstream of the bow shock
  (BS) of 92 quasi-perpendicular (left) and 40 quasi-parallel (right) bow shock
  crossings. Shown are from top to bottom the corrected electron
  density $N_e^*$, the density of energetic protons $N_{2p}$, the
  electron temperature $T_e$, and the proton temperature anisotropy
  $T_{p\perp}/T_{p\|}$.}
\end{figure}

\subsection{Plasma and magnetic field parameters}
The time series adjusted in the way described above are superimposed by
averaging 10-s bins. The averages are performed geometrically in order to
reduce the dominance of cases with large dynamical ranges.  The result
of the superposition is shown in Fig.~\ref{ppsp1} and
Fig.~\ref{ppsp2}. The averages of the downstream and
upstream values, excluding the 4 bins closest to the bow shock on both
sides, are given in Table~\ref{mittelwerte}, together with the
ratios of the downstream to the upstream values.

\begin{table*}
\renewcommand{\arraystretch}{1.5}
\tablewidth{20pc}
\caption{\label{mittelwerte} Averages of plasma and magnetic field
  parameters}\tablenum{1}

\begin{tabular}{llccc}
\tableline
& & Upstream & Downstream & Ratio \\
\hline
\hline
 & q-$\perp$ & $9.4\pm 0.1$  & $28.5\pm 0.8$  & $3.0\pm 0.1$ \\
 \raisebox{2.0ex}[-2.0ex]{$B$ [nT]}& q-$\|$ & $10.9\pm 0.8$ & $19.6\pm 1.1$
 & $1.8\pm 0.2$  \\ 
\hline
 & q-$\perp$ & $436\pm 6$  & $198\pm 7$  & $0.45\pm 0.03$ \\
\raisebox{2.0ex}[-2.0ex]{$v_p$ [km/s]} & q-$\|$ & $332\pm 16$ & $171\pm 17$
& $0.52\pm 0.08$  \\ \hline

 & q-$\perp$ &$-401\pm 9$  & $-122\pm 7$  & $0.30\pm 0.02$ \\
\raisebox{2.0ex}[-2.0ex]{$v_n$  [km/s]} & q-$\|$ &
$-283\pm 18$ & $-96\pm 16$ & $0.34\pm 0.08$ \\ \hline
 & q-$\perp$ &$0.6\pm 0.1$  & $2.3\pm 0.5$  & $4.0\pm 1.4$ \\ 
\raisebox{2.0ex}[-2.0ex]{$\delta B_2$ [nT]} & q-$\|$ & $1.9\pm 0.3$ & $3.2\pm 0.5$ & $1.7\pm0.5$  \\ \hline
 & q-$\perp$ & $7.4\pm 0.1$ & $20.1\pm
0.4$ & $2.7\pm 0.1$ \\ 
\raisebox{2.0ex}[-2.0ex]{$N_e^*$
[cm$^{-3}$]} & q-$\|$ & $7.8\pm 0.3$ & $14.4\pm 1.1$  & $1.8\pm 0.2$ \\ \hline
 & q-$\perp$ &$0.15\pm 0.01$ & $0.38 \pm 0.03$  &$ 2.6\pm 0.3$ \\
\raisebox{2.0ex}[-2.0ex]{$N_{2p}$ [$10^{-2}$ cm$^{-3}$]} & q-$\|$ & $4.0\pm
0.5$ & 2.6$\pm 0.5$ & $0.7\pm 0.2$  \\  \hline 
 & q-$\perp$ &  $27.5\pm 0.4$  & $71.3\pm
0.7$  & $2.6\pm 0.1$ \\ \raisebox{2.0ex}[-2.0ex]{$T_e$
  [$10^{4}$ K]}
 & q-$\|$ & $43\pm 1$ & $73\pm 3$ & $1.7\pm0.1$ \\ 
\hline
 & q-$\perp$ &   & $1.51\pm 0.06$ & \\
\raisebox{2.0ex}[-2.0ex]{$T_{p\perp}/T_{p\|}$} & q-$\|$ &
\raisebox{2.0ex}[-2.0ex]{---}& $1.05\pm 0.03$ &
\raisebox{2.0ex}[-2.0ex]{---} \\  \hline
\end{tabular}
\caption{Averages of 92 quasi-perpendicular (q-$\perp$) and 40
  quasi-parallel (q-$\|$) bow shock crossings from 120 to 20 seconds upstream and 20 to 240
  seconds downstream: the magnitude $B$ of the magnetic field, the
  magnitude $v_p$ of the
  proton velocity, the proton velocity $v_n$ parallel to the bow
  shock normal vector, the magnetic fluctuations $\delta B_2$, the
  corrected electron density $N_e^*$, the density of energetic protons
  $N_{2p}$, the electron temperature $T_e$, and the proton temperature
  anisotropy $T_{p\perp}/T_{p\|}$. Ratios of the downstream to the
  upstream values are given in the last column.}
\end{table*}

The top panel of Fig.~\ref{ppsp1} shows the magnitude $B$ of the
magnetic field. It increases steeply by a factor of 3 at the
quasi-perpendicular bow shock and gradually by a factor of 1.8 at the
quasi-parallel bow shock. The proton bulk velocity $v_p$, shown in the
next panel, decreases to somewhat less than half of its solar wind
value for the quasi-perpendicular bow shock and to slightly more than
half for the quasi-parallel bow shock. The third panel shows the
proton velocity parallel to the shock normal vector.  The latter is
calculated from the Fairfield bow shock model [{\it Fairfield}, 1971].
For both categories $v_n$ decreases by more than the magnitude of the
proton velocity $v_p$, indicating that the plasma is deflected away
from the bow shock normal direction in order to flow around the
magnetopause.  In the last panel we show the root mean square
amplitude $\delta B_2$ of the high resolution magnetic field
measurements during one spin period:
\begin{equation}
  \label{edeltab}
  \delta B_2=\left[\sum_{k=1}^3 \frac{1}{n}\sum_{i=1}^n(B_{k,i}-\bar{B}_k)^2 \right]^{1/2}
\end{equation}
where $i=1,..., n$ counts the measurements during one spin period, and
$B_k$, $k=1,2,3$ denote the magnetic field vector components.
$\bar{B}_k$ is the average of $B_k$ taken during one spin period. At
the keytime, $\delta B_2$ is approximately half the jump of the
magnitude of the magnetic field. Further upstream and downstream it is
a measure of wave activity with Doppler-shifted periods shorter than
the spin period of about 4.3 s.  Although the magnetic field increases
only gradually at the quasi-parallel bow shock, $\delta B_2$ is strongly
enhanced at the bow shock crossing time and has a maximum immediately
downstream of the shock. This jump in $\delta B_2$, as a parameter
measured independently of the velocity, is a confirmation that the
selection of the crossing times with the help of the velocity jump
(Section 2) is reasonable. In the foreshock region upstream of the
quasi-parallel bow shock, $\delta B_2$ is considerably higher than in
the solar wind regime upstream of the quasi-perpendicular bow shock.

The first panel of Fig.~\ref{ppsp2} shows the electron density $N_e^*$,
approximately corrected for the low-energy cut-off of the plasma
instrument [{\it Sckopke et al.}, 1990]. Like the magnetic field, the
density rises sharply at the quasi-perpendicular bow shock, whereas it
increases gradually at the quasi-parallel bow shock. The second panel
shows $N_{2p}$, the proton density in the
energy interval 8-40 keV. $N_{2p}$ increases by a factor of 2.6 at the
quasi-perpendicular bow shock. In the foreshock region upstream of the
quasi-parallel bow shock $N_{2p}$ is an order of magnitude higher than
upstream of the quasi-perpendicular bow shock and decreases by a
factor of about 0.7 in the downstream region. 

In the next panel the
electron temperature $T_e$ is shown. Downstream of the quasi-parallel bow
shock crossing it has about the same value as downstream of the
quasi-perpendicular bow shock. However, upstream of the
quasi-parallel bow shock $T_e$ is about a factor of 1.6 higher than upstream
of the quasi-perpendicular bow shock. The reason for this is again the
foreshock region where the solar wind kinetic energy is already partly
transformed into thermal energy. The last panel shows the proton
temperature anisotropy $T_{p\perp}/T_{p\|}$.  The plasma instrument
did not resolve the cold, supersonic distributions of the solar wind
ions. The calculated proton densities and temperatures are therefore
not reliable in the solar wind regime.  Hence, the proton temperature
anisotropy cannot be determined in the upstream region and is
therefore set to 1. Whereas the proton temperature anisotropy
downstream of the quasi-parallel bow shock is insignificant, there is
a strong   proton temperature anisotropy, $T_{p\perp}/T_{p\|} > 1.4$,
downstream of the quasi-perpendicular bow shock, with a maximum value of
more than 2 immediately behind the shock.
Comparing the downstream values of the electron and proton
temperatures (not shown), we find 
$T_p \approx 4.3 \times 10^6\,{\rm K} \approx 6\,T_e$ 
for the quasi-parallel bow shock and
$T_p \approx 2.9 \times 10^6\,{\rm K} \approx 4\,T_e$
for the quasi-perpendicular bow shock. 
Whereas the electron temperature is slightly anisotropic, 
$T_{e\perp}/T_{e\|}\approx 0.9$, upstream and downstream of quasi-perpendicular
shocks, no significant anisotropy is observed at quasi-parallel shocks.

\begin{figure*}[h!]
\epsscale{0.7}
\plotone{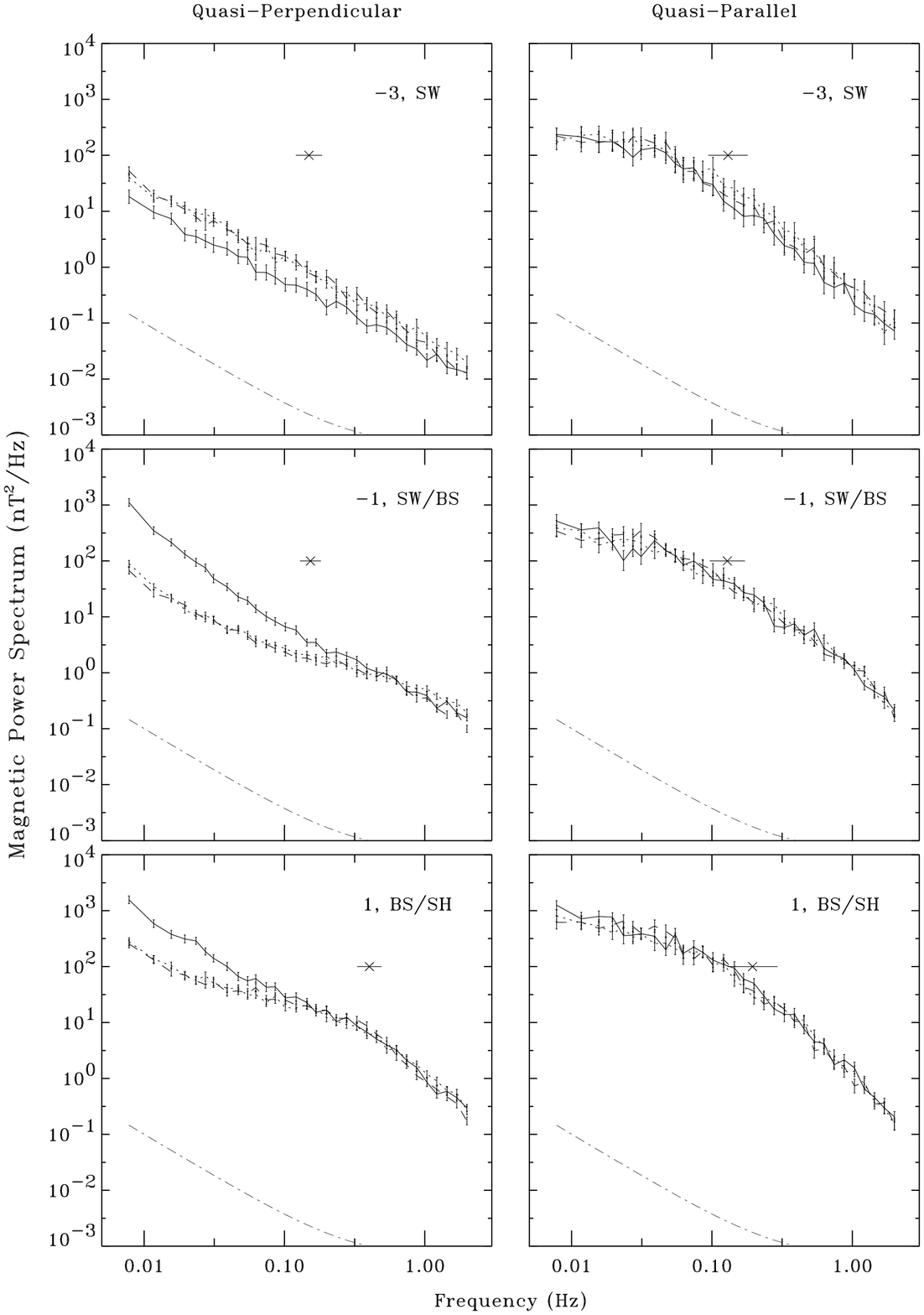}
\caption{\label{flucsp1} Superposed-epoch analysis of magnetic spectra
  of 92 quasi-perpendicular (left) and 40
  quasi-parallel bow shock crossings (right) from 3\,min upstream to 1 min
  downstream. Solid line: compressive component, dashed line: left
  hand polarized component, dotted line: right hand polarized
  component. The cross with the horizontal  error bar marks the proton 
  cyclotron frequency. The acronyms SW, BS, SH mean solar wind, bow
  shock, and magnetosheath, respectively. } 
\end{figure*}

\begin{figure}[h!]
\epsscale{0.7}
\plotone{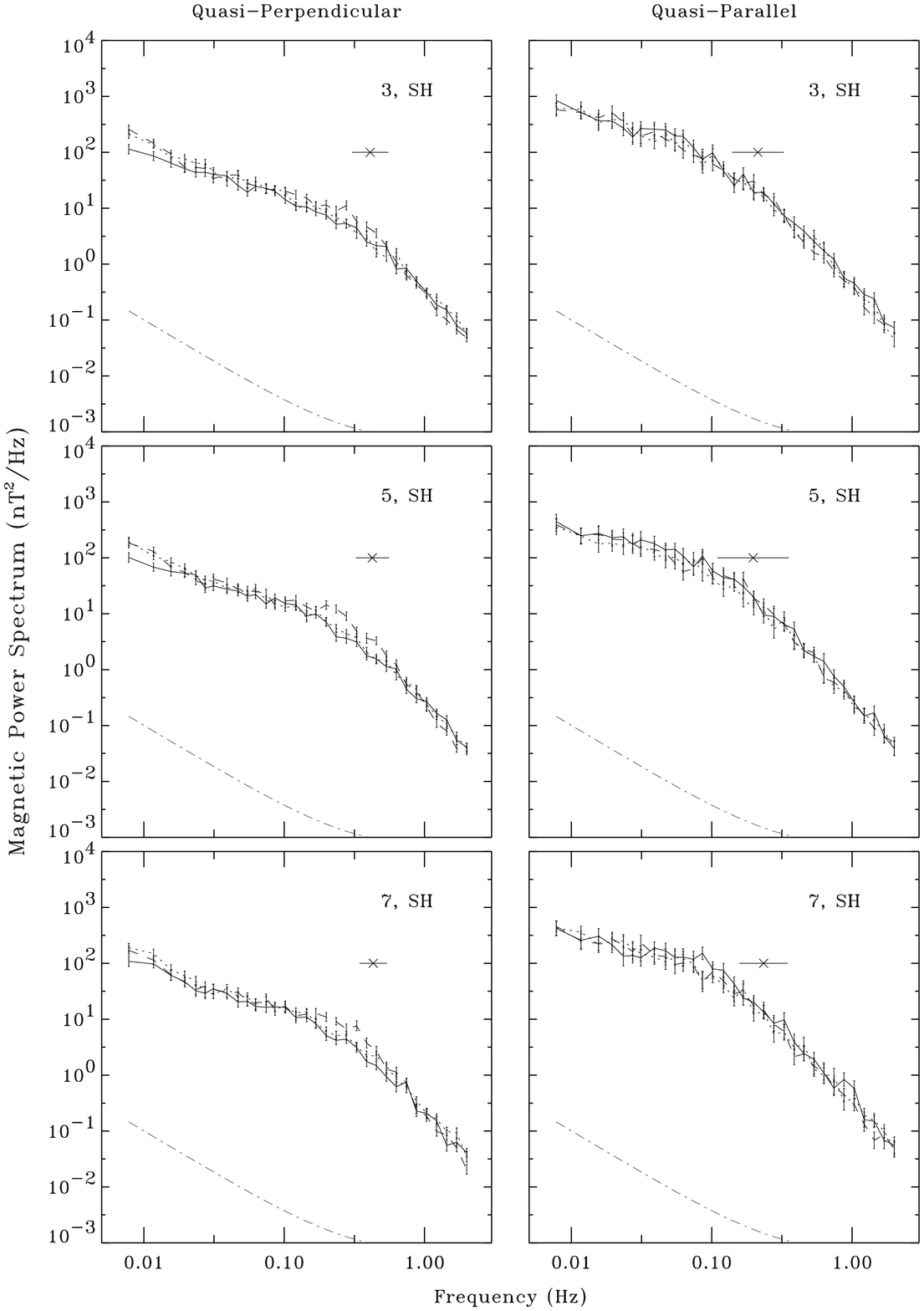}
\caption{\label{flucsp2} Superposed-epoch analysis of magnetic spectra
  of 92 quasi-perpendicular (left) and 40
  quasi-parallel bow shock crossings (right) from 3\,min to 7\,min
  downstream. Same format as Fig.~\ref{flucsp1}. } 
\end{figure}

\subsection{Low frequency magnetic fluctuations}
In order to analyze the low frequency magnetic fluctuations we perform
a spectral analysis of the magnetic field using a cosine-bell filter
[see, e.g., {\it Bauer et al.}, 1995]. The Fourier transform is taken
over a time interval of 4 min. Fig.~\ref{flucsp1} and
Fig.~\ref{flucsp2} show the resulting power spectra of the compressive
and the right- and left-hand polarized modes, respectively. In each
graph the center time (-3,-1,1,3,5,7) of the transformed time interval
is given in minutes relative to crossing time. A cross marks the
proton cyclotron frequency
$f_{cp}=eB/2\pi m_p$
with $m_p$ the proton mass.  The solar wind (SW) spectrum of
Fig.~\ref{flucsp1}, 3\,min upstream of the quasi-perpendicular bow
shock, shows a structureless decrease to higher frequencies following
the power law $S \sim f^{-1.3}$. The compressive mode lies below the
transverse modes that represent Alfv\'en waves frequently observed in
the interplanetary medium. They are thought to have their origin in
the vicinity of the Sun [{\it Belcher \& Davis}, 1971]. The spectrum
upstream of the quasi-parallel bow shock has much more power than that
upstream of the quasi-perpendicular bow shock. It also has a different
structure: For lower frequencies it shows a flatter decrease ($S \sim
f^{-0.5}$), while for higher frequencies it decreases more steeply ($S
\sim f^{-2.0}$). The kink in the spectrum lies below the proton
cyclotron frequency.

The next power spectra (-1, SW/BS) contain magnetic field data from
the upstream region and the bow shock itself. At the
quasi-perpendicular bow shock the compressive mode at low
frequencies is one order of magnitude higher compared to the transverse
modes and follows a power law of $S \sim f^{-2.1}$. This represents
simply the spectrum of the jump of the magnetic field across the shock
filtered with a cosine bell function.  The spectra of the transverse
modes are a little higher than 2\,min earlier and the decrease with
frequency is not constant any more. At the quasi-parallel bow shock
the increase of the magnetic field is not visible, which is not
surprising since the magnetic field increases only gradually. Level
and structure of the spectrum are similar to those calculated 2 min
earlier.

The spectra (1, SW/BS) contain magnetic field data from again the bow
shock itself and from the magnetosheath just downstream of the bow
shock. At the quasi-perpendicular bow shock the compressive mode
behaves similar to the spectrum 2\,min earlier, whereas the spectral
power of the transverse modes is higher than 2\,min earlier. Just below
the proton cyclotron frequency first indications of a plateau are
visible. At the quasi-parallel bow shock wave activity is
significantly enhanced by a factor of 4 compared to the
upstream spectra. Again all three modes behave similar. The proton
cyclotron frequencies increase according to the magnetic field
increase by a factor of 3 at the quasi-perpendicular bow shock and by
a factor of 2 at the quasi-parallel bow shock.

Figure~\ref{flucsp2} shows that for both categories of shock crossings
the spectra do not change much in the interval 2 to 8\,min downstream
of the bow shock. Below $f_{cp}$ the compressive and the right-hand polarized
modes downstream of the quasi-perpendicular bow shock follow a power
law $S \sim f^{-1.1}$, whereas the spectral energy of the left-hand
polarized mode is clearly enhanced in the frequency interval from
about 0.1\,Hz to about the proton cyclotron frequency. This fact is
investigated more carefully in Section 4. Downstream of the
quasi-parallel bow shock the spectral energy is again higher than
downstream of the quasi-perpendicular bow shock. For $f<f_{cp}$ it
follows the power law $S\sim f^{-0.8}$. 3\,min after the crossing 
the spectral energy is higher than in the 2 later spectra but already
lower than directly at the bow shock. For both categories the spectral
energy decreases steeply ($S\sim f^{-2.6}$) for $f>f_{cp}$.

\subsection{Discussion}

The processes in the bow shock transition region itself cannot be
described in terms of magnetohydrodynamics (MHD). However, if one considers
the bow shock as an infinitesimally thin discontinuity one can derive
from the MHD equations the Rankine-Hugoniot jump conditions (see,
e.g., {\it Siscoe}, 1983), which are relations for the conditions in
the plasma upstream and downstream of the discontinuity under the
assumption of time independence.

For mass continuity the jump condition is
\begin{equation}
  \label{con}
 n_2 v_{n2} - n_1 v_{n1} = 0  
\end{equation}
where the subscripts 1 and 2 denote the upstream and downstream values
of the corresponding parameters, respectively.  According to
Eq.~(\ref{con}) the product of the downstream-to-upstream ratio of the
electron density $N_e$ and the normal proton velocity $v_n$, which is
equivalent to the plasma bulk speed normal to the discontinuity, must
be unity.  For the quasi-parallel events this product is observed to
be about 0.6.  This fact is not surprising, since the considered
upstream time profiles are not taken from the quiet solar wind regime
but from the dynamic foreshock region, which is highly time dependent.
For the quasi-perpendicular events this product is observed to be
about 0.8, which means that the jump condition Eq.~(\ref{con}) is not
fully satisfied. This could be explained by the fact that we have not
measured the exact electron density, but only the density of electrons
with energies between 15 eV and 30 keV. Electrons with higher energies
are negligible since already in the energy range of 1.8 - 30 keV the
upstream density is of the order of $10^{-5}$ cm$^{-3}$. However,
particularly in the cold solar wind, electrons with energies below the
instrument cut-off contribute an essential part to the total electron
density. Therefore we use the corrected electron density $N_e^*$. The
correction is calculated with the assumption of a Maxwellian
distribution of the measured temperature.  However, as has been
measured, e.g., recently by the Wind spacecraft [Fig.~4 of {\it Lin},
1997], the quiet solar wind flow cannot be described by a single
Maxwellian distribution over its whole energy range. Therefore the
electron density could easily be overestimated by the correction,
especially in the cold solar wind regime where the low energy
electrons are more important than in the warmer magnetosheath regime.
In order to satisfy the jump condition Eq.~(\ref{con}) we can estimate,
that the value of the corrected electron density upstream of the
quasi-perpendicular bow shock is about 20\% too high if we assume that
the error in the downstream value is not of importance. Other factors
for the apparent deviation from mass continuity could be the unknown
bow shock motion, which changes the downstream plasma velocity by a
higher percentage than the upstream plasma velocity, and the
uncertainty of the shock normal vector.

For an exactly parallel shock wave one derives from the
Rankine-Hugoniot jump condition that the magnetic field remains
constant in magnitude and direction and one is left with the equations
for a purely hydrodynamic shock wave.  In the case of an exactly
perpendicular shock wave the jump conditions give the relation
\begin{equation}
  \label{esenk}
\frac{n_2}{n_1}=\frac{B_2}{B_1}=\frac{v_1}{v_2}  
\end{equation}

For the limit of high Mach numbers, i.e., the Alfv\'en Mach number
$M_A \rightarrow \infty$ and the sonic Mach number $M_{s} \rightarrow
\infty$, Eq.~(\ref{esenk}) has a value of 4 for $\gamma=5/3$. Since we do
not observe the extreme cases of exactly perpendicular and exactly
parallel geometry but quasi-perpendicular and quasi-parallel shock
wave crossings with finite Mach numbers the numbers given above are
not reached. For the quasi-perpendicular bow shock the ratio
Eq.~(\ref{esenk}) is about 3 for the average magnetic field $B$ and proton
normal velocity $v_n$, and somewhat less for the corrected electron
density $N_e^*$ according to the above mentioned probable
overestimation of this parameter in the solar wind regime.  For the
quasi-parallel bow shock the change in the magnitude of the magnetic
field is significantly lower than for the quasi-perpendicular bow
shock, which is consistent with theory.

In their study of subcritical quasi-perpendicular shocks,
{\it Thomsen et al.}, [1985] found that the downstream electron temperature is
nearly isotropic, $T_{e\perp 2}/T_{e\| 2}\approx 0.9$ , 
and that the downstream-to-upstream ratio 
$T_{e\perp 2}/T_{e\perp 1}$ of the perpendicular temperature 
is approximately equal to the ratio $B_2/B_1$ of the magnetic field strength. 
From the latter result they concluded that the net heating is adiabatic, 
although $T_{e\perp}/B$ is not constant. 
Obtaining averages, $T_{e\perp 2}/T_{e\| 2}\approx 0.9$,
$T_{e\perp 2}/T_{e\perp 1}\approx 2.6$, and $B_2/B_1\approx 3.0$, we can
confirm these results for our data set of (subcritical and supercritical)
quasi-perpendicular shocks. Moreover, isotropy of the downstream electron 
temperature and adiabatic net heating is also found for our data set of
quasi-parallel shocks, for which we obtain averages
$T_{e\perp 2}/T_{e\| 2}\approx 1.0$,
$T_{e\perp 2}/T_{e\perp 1}\approx 1.7$, and $B_2/B_1\approx 1.8$.

An interesting question of shock physics is how the dissipated bulk flow energy
of the solar wind is partitioned amongst ion and electron heating. For the 
average proton-to-electron temperature ratio we found $T_p/T_e\approx 6$
downstream of the quasi-parallel bow shock and $T_p/T_e\approx 4$
downstream of the quasi-perpendicular bow shock. Hence, 
for quasi-parallel shocks proton heating is more favored with respect to
electron heating than for quasi-perpendicular shocks.

Let us turn to the low frequency waves of Fig.~\ref{flucsp1} and
Fig.~\ref{flucsp2}. 
At quasi-parallel bow shocks the wave power observed 3\,min upstream of the 
keytime is much higher than the power of the interplanetary Alfv\'en waves 
observed upstream of quasi-perpendicular shocks. 
This enhanced power reflects upstream waves generated in the foreshock. 
Observations of upstream waves have recently been reviewed by 
{\it Greenstadt et al.} [1995] and {\it Russell \& Farris} [1995]. 
The nonlinear steepening of the shock leads to
whistler precursors phase standing in the shock frame. The interaction between
ions reflected at the shock and the incoming solar wind can drive ion beam
instabilities. These are probably the source of large-amplitude waves
observed at periods around 30\,s.
Finally, there are upstream propagating whistlers with frequencies around
1\,Hz, which seem to be generated directly at the shock.
The most striking feature in the average spectrum of upstream waves observed
3\,min before the keytime is the kink at $0.04\,{\rm Hz}\approx f_{cp}/3$. 
The average power measured in the flat portion 0.01--0.04\,Hz of the spectrum
corresponds to a mean square amplitude $\delta B_1\approx 4\,nT$ or 
$\delta B_1/B\approx 0.4$. Large-amplitude waves observed in this frequency 
range [{\it Le \& Russell}, 1992; {\it Blanco-Cano \& Schwartz}, 1995] 
have been interpreted 
as upstream propagating magnetosonic waves excited by the right-hand resonant 
ion beam instability, upstream propagating Alfv\'en/ion cyclotron waves excited 
by the left-hand resonant ion beam instability, or downstream propagating 
magnetosonic waves excited by the non-resonant instability. Whereas
the upstream propagating magnetosonic waves should be left-hand polarized in
the shock frame (and also in the spacecraft frame), the other two wave types
should be right-hand polarized. This might explain why none of the two circular
polarizations dominates in our average spectra. Moreover, the compressional
component is comparable to the two transverse components. This shows that the
waves propagate at oblique angles to the magnetic field. For oblique
propagation, low frequency waves have only a small helicity [{\it Gary}, 1986].
Thus they are rather linearly than circularly polarized. 

The power spectra presented by {\it Le \& Russell} [1992] exhibit clear peaks 
at $f\approx f_{cp}/3$. Looking into the spectra of individual time intervals,
we find that sometimes the IRM data exhibit similar spectral peaks. However, 
most of the individual spectra do not have clear peaks, but are rather flat in
the range 0.01--0.04\,Hz like the average spectrum of Fig.~\ref{flucsp1}.
The steep decrease of the power above about $f\approx f_{cp}/3$ is common to
our spectra and those reported previously. In fact, the maximum growth rate of
the ion beam generated waves is expected for frequencies below the proton
cyclotron frequency [e.g., {\it Scholer et al.}, 1997].

In Fig.~\ref{ppsp1} we saw that magnetic fluctuations above 0.23\,Hz 
in the foreshock have root mean
square amplitudes $\delta B_2\approx 1.5$\,nT. This is comparable to typical
amplitudes of the upstream propagating whistlers. In individual time intervals 
these narrow-band waves lead to clear spectral peaks at frequencies 
around 1\,Hz. However, since the frequency varies from event to event, 
no such peak appears in the average spectra.

At the keytime of quasi-parallel bow shocks we observed a clear enhancement of
the wave power. This enhancement can either be due to an amplification of 
the upstream waves or due to wave generation at the shock interface. 
Wave generation at the shock due to the interface instability has been found in
hybrid simulations of {\it Winske et al.} [1990] 
and {\it Scholer et al.} [1997]. This instability is driven by the 
interaction between the incoming solar wind ions and the heated downstream 
plasma at the shock interface. 
Amplification of upstream waves has been predicted by 
{\it McKenzie \& Westphal} [1969] 
who analyzed the transmission of MHD waves across a fast shock.
They found that the amplitude of Alfv\'en waves increases by a factor of 3. 
For compressional waves the amplification can even be stronger. However, the
hybrid simulations of {\it Krauss-Varban} [1995] show that the transmission 
of waves across the shock is complicated by mode conversion.

The proton temperature anisotropy downstream of the
quasi-perpendicular bow shock serves as a source of free energy.
According to both observations and simulations this kind of free
energy drives two modes of low frequency waves under the plasma
conditions in the magnetosheath: the ion cyclotron wave and the mirror
mode (see e.g., {\it Sckopke et al.}, 1990, {\it Hubert et al.}, 1989
and {\it Anderson et al.}, 1994 for observations, {\it Price et al.},
1986, and {\it Gary et al.}, 1993 for simulations, and {\it Schwartz
  et al.} 1996 for a review). 

Which of these waves grow under which conditions is investigated in Section 4, 
where we divide the crossings of quasi-perpendicular shocks into cases with 
low and high upstream $\beta$, respectively.
It turned out that for this classification by $\beta$ the differences become
somewhat clearer than for a classification by upstream Mach number.
The critical Mach number $M^*$ above which ion reflection is required to
provide the necessary dissipation is
strongly dependent on the plasma-$\beta$ [{\it Edmiston \& Kennel}, 1984]. 
We have calculated the ratio $M_{ms} / M^*$ for our shock
crossings and have found that all subcritical shocks are
low-$\beta$, i.e., that the classification low-$\beta$ versus
high-$\beta$ is more or less identical to the classification
subcritical versus supercritical. The reason for this is that 
$M^* \rightarrow 1$ for $\beta \gg 1$. 
As the excitation of mirror and ion cyclotron waves depends on $\beta$, 
the results of Section~4 should be interpreted
as the effect of the plasma-$\beta$ and not as an effect of
subcritical or supercritical Mach numbers. 

For the quasi-parallel bow shocks, we could not investigate the difference
between subcritical and supercritical shocks, because no subcritical 
quasi-parallel shock was identified in the data set. Trying higher thresholds
for the division into low and high Mach numbers, we did not find any 
qualitative differences. In this context it should be noted that only one of
the cases in our data set has an Alfv\'en Mach number in the range 
$M_A\le 2.3$, for which quasi-parallel shocks are steady according to the
hybrid simulations of {\it Krauss-Varban \& Omidi} [1991].

 \section{Comparison of quasi-perpendicular low-$\beta$ and
  high-$\beta$ bow shock crossings}

In order to reveal the origin of the left-hand polarized component in
the power spectra downstream of the quasi-perpendicular bow shock
(Fig.~\ref{flucsp2}) we divide the crossings into classes with low
($< 0.5$) and high ($> 1.0$) upstream $\beta$ and compare these two
classes. There are 20 low-$\beta$ and 47 high-$\beta$ cases. The
crossings with $0.5 \le \beta \le 1.0$ are not included in this
analysis in order to emphasize the differences between the low-$\beta$
and high-$\beta$ regimes.

\begin{figure}[h!]
\epsscale{0.75}
\plotone{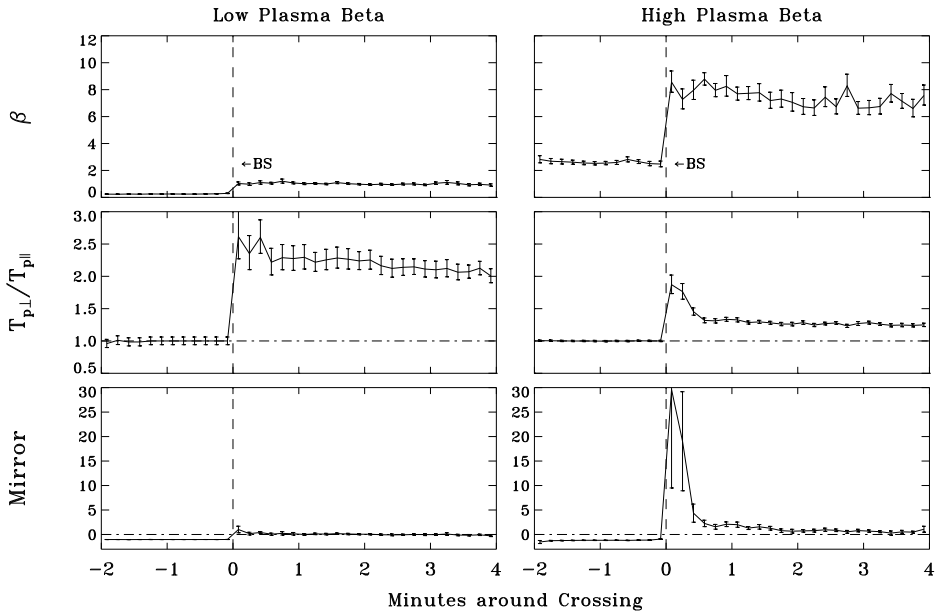}
\caption{\label{pplh}Superposed-epoch analysis of the plasma parameter
  $\beta$, the proton temperature anisotropy $T_{p\perp}/T_{p\|}$, and
  the mirror instability criterion from 2\,min upstream to 4 min
  downstream of the quasi-perpendicular bow shock (BS) on the left for
  20 low-$\beta$, on the right of 47 high-$\beta$ crossings.} 
\end{figure}

\subsection{Plasma and magnetic field parameters}

In Fig.~\ref{pplh} we show some interesting differences between
the low-$\beta$ and high-$\beta$ categories.  Of course the plasma
parameter $\beta$ differs essentially.  In the upstream region $\beta$
is derived by setting the proton density to the corrected electron
density and the proton temperature to $10^{-5}$ K, the long term
average of the proton temperature, since proton distribution functions
are not well measured in the cold solar wind (Section 3.1).  The
classification for the low-$\beta$ and high-$\beta$ categories is
derived from the estimated plasma $\beta$ in the upstream region. The
first panel of Fig.~\ref{pplh} shows that the same classification
could be obtained using the plasma-$\beta$ of the downstream region
with the limits shifted to larger values.  The most striking
differences between the low-$\beta$ and high-$\beta$ bow shock are
shown in the next two panels, i.e., the proton temperature anisotropy
$T_{p\perp}/T_{p\|}$ and the mirror wave instability criterion.  Both
parameters are again determined only in the downstream region.  The
instability criterion for almost perpendicular propagation of the
mirror mode in its general form [{\it Hasegawa}, 1969] is given by
\begin{equation}  \label{mirin} -1 + \sum_j  \beta_{j\perp} \left ( 
\frac{\beta_{j\perp}} {\beta_{j\|}}-1\right )    > 0.  
\end{equation} 
The subscript $j$ denotes the particle species ($j$ = $e$, $p$ for
electrons and protons, respectively).

Downstream of the quasi-perpendicular low-$\beta$ bow shock the proton
temperature anisotropy is very high, $T_{p\perp}/T_{p\|}\approx 2.5$,
immediately behind the shock and remains high, $T_{p\perp}/T_{p\|} >
2$, throughout the whole magnetosheath interval investigated. Downstream
of the quasi-perpendicular high-$\beta$ bow shock the proton
temperature anisotropy is also significant, but lower compared to the
low-$\beta$ bow shock, i.e., $T_{p\perp}/T_{p\|}\approx 1.8$ just
behind the bow shock and $T_{p\perp}/T_{p\|}\approx 1.3$ further
downstream. The mirror instability criterion is only marginally satisfied
immediately downstream of the quasi-perpendicular low-$\beta$ bow shock
and is not satisfied at later times, since the instability criterion does
not only depend on the particle temperature anisotropy but also on the
absolute value of $\beta$. Downstream of the quasi-perpendicular
high-$\beta$ bow shock the mirror instability criterion is 
satisfied in the entire interval of 4\,min behind the shock. Extremely
high values of the left-hand side of Eq.~(\ref{mirin}) are occasionally
observed immediately behind the shock.

\begin{figure}[h!]
\epsscale{0.7}
\plotone{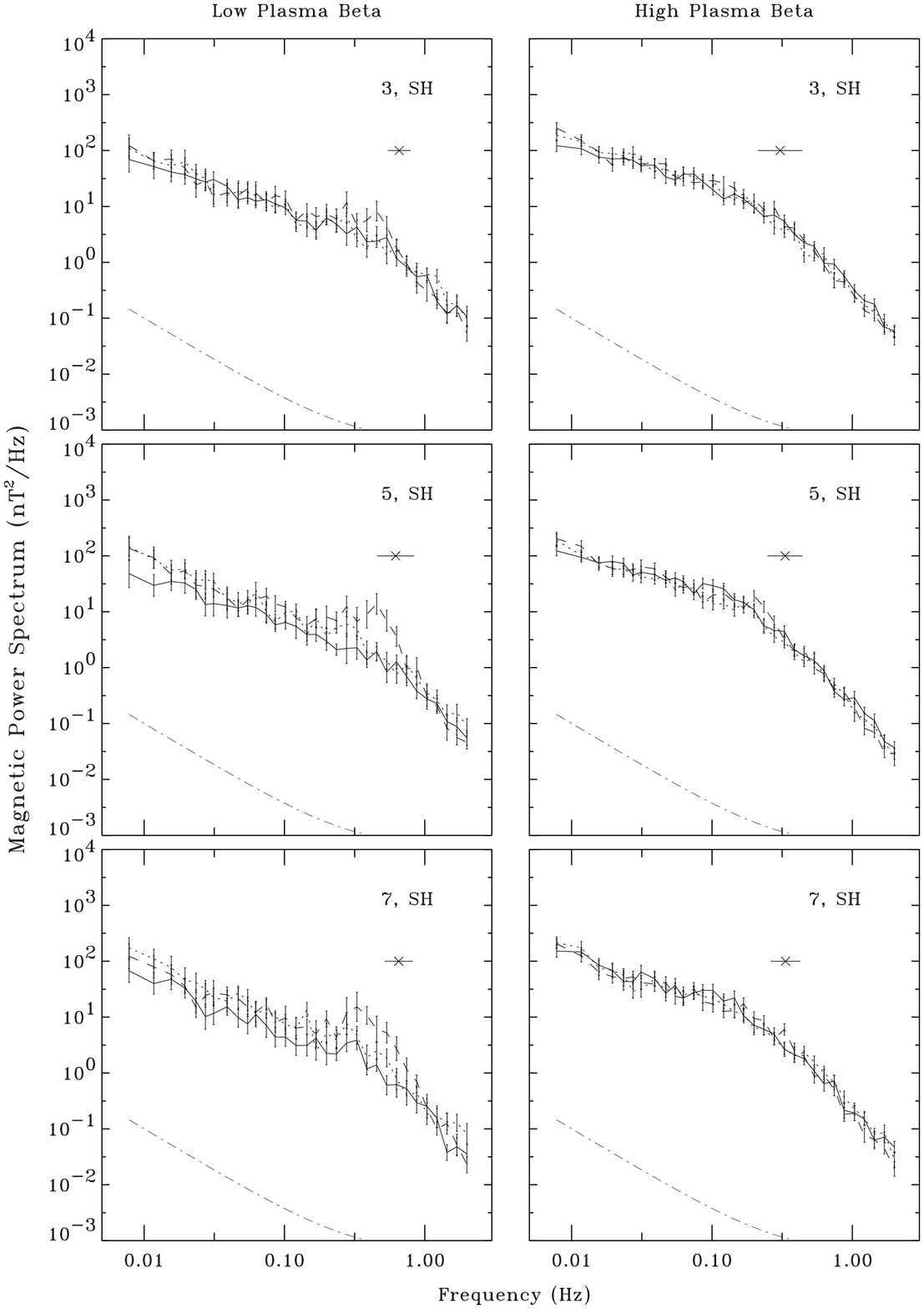}
\caption{\label{fluclh} Superposed-epoch analysis of magnetic power spectra
downstream of 20 quasi-perpendicular low-$\beta$ and 47 quasi-perpendicular
high-$\beta$  bow shock crossings. Same format as Fig.~\ref{flucsp2}.}
\end{figure}

\subsection{Low frequency magnetic fluctuations}
Fig.~\ref{fluclh} shows the magnetic power spectra downstream of the
quasi-perpendicular low-$\beta$ (left) and high-$\beta$ (right) bow
shock, 3, 5, and 7\,min after the crossing time. Now it becomes
obvious that the left-hand polarized mode dominates only behind the
quasi-perpendicular bow shock with low-$\beta$ in a frequency interval
below the proton cyclotron frequency. 5 and 7 min downstream, the
left-hand polarized mode has up to one order of magnitude more power
spectral density than the compressive and the right-hand polarized
components.  Below the frequency range where the left-hand polarized
component dominates in the low-$\beta$ cases, the spectrum downstream
of the quasi-perpendicular high-$\beta$ bow shock shows a weaker
gradient than downstream of the low-$\beta$ bow shock. In addition,
the compressive mode lies on the same level as the transverse modes
and occasionally at somewhat higher levels, whereas 5 and 7\,min
downstream of the low-$\beta$ bow shock the compressive mode lies
clearly below the transverse modes.Due to the higher magnetic field
the proton cyclotron frequency is a factor of about 2.3 higher
downstream of the low-$\beta$ bow shock ($B \approx$ 45 nT) than
downstream of the high-$\beta$ bow shock ($B \approx$ 23 nT).

At this point, let us collect some numbers for typical wave amplitudes. 
For that purpose we use the root mean square amplitude $\delta B_2$ 
given in Eq.~(\ref{edeltab}), which has already been shown in Fig.~\ref{ppsp1} 
and can be obtained by integrating the power spectra for frequencies 
above 0.23\,Hz, and we use the root mean square amplitude $\delta B_1$ 
of fluctuations in the frequency range 0.01--0.04\,Hz. 
The range 0.01--0.04\,Hz has been chosen, because it corresponds to 
the flat portion of the spectrum observed 3\,min upstream of the keytime 
of quasi-parallel bow shocks.
These upstream waves have $\delta B_1\approx 4$\,nT or 
$\delta B_1/B\approx 0.4$. In contrast, 
the Alfv\'en waves seen in the solar wind upstream of quasi-perpendicular bow 
shocks have $\delta B_1\approx 0.8$\,nT or $\delta B_1/B\approx 0.09$. 
At the keytime of quasi-parallel bow shocks we observed a clear enhancement of
the wave power, which leads to 
$\delta B\approx 8$\,nT or $\delta B/B\approx 0.5$. 

In this section we saw that the characteristics of the wave activity downstream 
of quasi-perpendicular shocks depend on $\beta$. For low $\beta$, the magnetic 
fluctuations above 0.23\,Hz are dominated by left-hand polarized fluctuations 
with $\delta B_2\approx 3$\,nT or $\delta B_2/B\approx 0.08$, which will be 
interpreted as ion cyclotron waves. For high $\beta$, typical amplitudes are 
$\delta B_2\approx 1.5$\,nT or $\delta B_2/B\approx 0.08$. 

\subsection{Discussion}

{\it Sckopke et al.} [1990] have performed a case study of low-$\beta$
subcritical bow shock crossings, using the AMPTE/IRM data of September
5, and November 2, 1984. These events are also included in our
quasi-perpendicular low-$\beta$ data set: There are 8 events
from September 5 and 3 events from November 2, 1984. {\it Sckopke et
  al.} [1990] identified the dominating left-hand polarized component
with the ion cyclotron wave which can be generated by a proton
temperature anisotropy (e.g., {\it Hasegawa}, 1975). The growth rate
of the ion cyclotron wave is positive when the instability criterion
is satisfied. This holds for the resonance with protons when 
\begin{equation}
  \label{eion}
  \frac{T_{p\perp}}{T_{p\|}}> \frac{f_{cp}}{f_{cp}-f}.
\end{equation}
The AMPTE/IRM plasma instrument did not resolve ion masses, therefore
all ions are assumed to be protons.

\begin{figure}[h!]
\epsscale{0.6}
\plotone{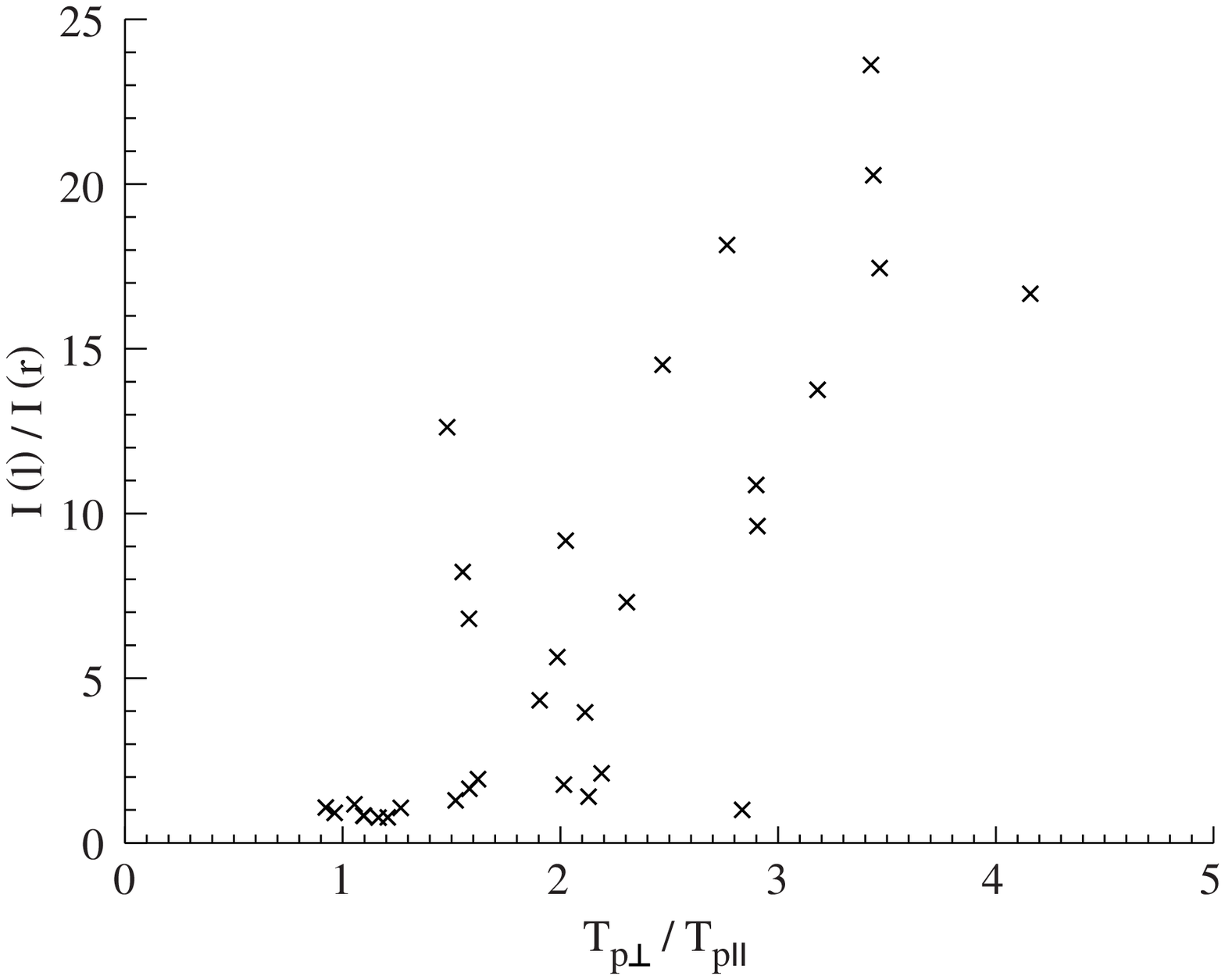}
\caption{\label{corrl} Ratio of the left-hand polarized to the right
  hand polarized component  $I(l)/I(r)$ against the proton temperature
  anisotropy $T_{p\perp}/T_{p\|}$
  ot 32 2-min intervals from 4 to 8\,min downstream of the
  quasi-perpendicular  low-$\beta$ bow shock.}
\end{figure}

 Figure~\ref{corrl} shows the ratio of the left-hand
 polarized to the right-hand polarized component in the frequency
 band 0.3--$0.8 f_{cp}$
 for 32 2-min intervals from 4 to 8\,min downstream of the
 quasi-perpendicular low-$\beta$ bow shock as a function of the proton
 temperature anisotropy. There is a clear correlation found between
 these two ratios with a correlation coefficient of 0.8. This shows
 that the wave intensity more than 4\,min downstream of the
 quasi-perpendicular bow shock depends strongly on the local
 temperature anisotropy.

\begin{figure}[h!]
\epsscale{0.6}
\plotone{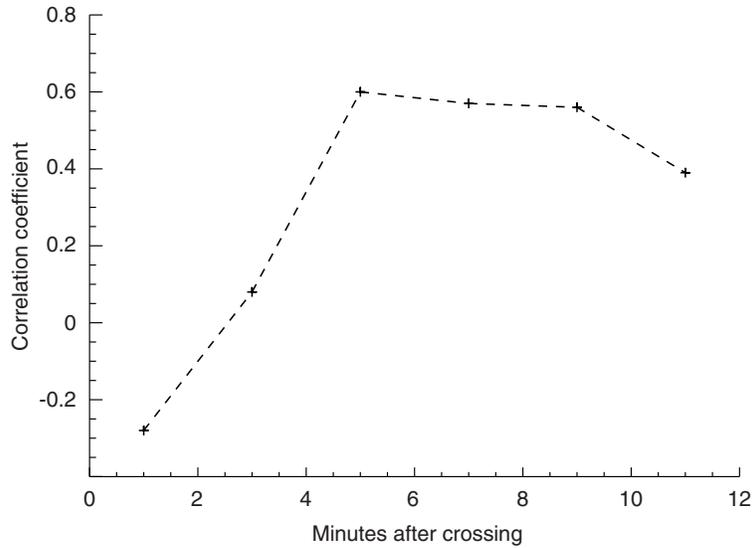}
\caption{\label{corrver} Temporal evolution of the correlation
  coefficient of the ratios $I(l)/I(r)$ and $T_{p\perp}/T_{p\|}$
  downstream of the  quasi-perpendicular  low-$\beta$ bow shock.}
\end{figure}

 The temporal evolution of the correlation coefficients, calculated
 for 2-min intervals downstream of the quasi-perpendicular low-$\beta$
 bow shock, is shown in Fig.~\ref{corrver}. The value of
 the correlation coefficient 11 minutes downstream is not reliable
 since only a limited data set extends so far downstream. Although the
 temperature anisotropy is highest immediately downstream of the bow
 shock, the best correlation is found around 5\,min downstream. This
 shows that the ion cyclotron waves need a certain time to develop in
 the moving plasma.
 
\begin{figure}[h!]
\epsscale{0.6}
\plotone{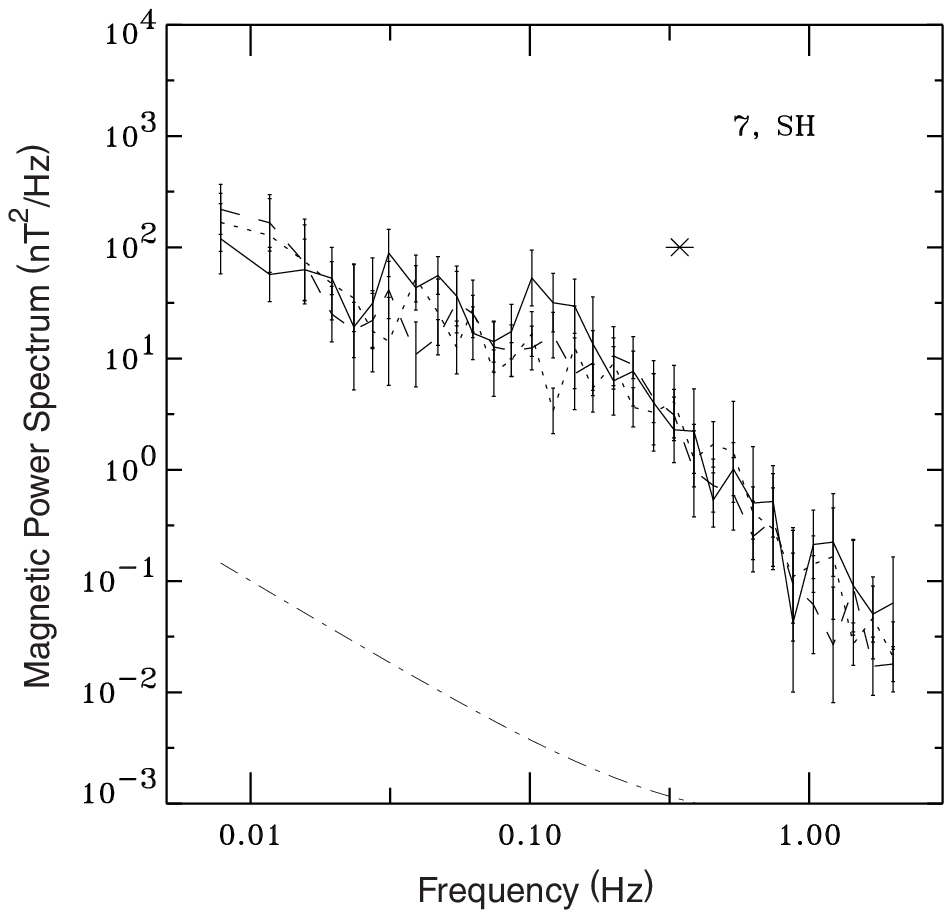}
\caption{\label{mmir} Superposed-epoch analysis of magnetic spectra
7\,min downstream of  9 quasi-perpendicular
high-$\beta$ bow shock crossings of which the mirror instability
criterion is well satisfied.}
\end{figure}

 Since downstream of the quasi-perpendicular high-$\beta$ bow shock
 the mirror instability criterion is satisfied on average, we have
 looked more carefully for this highly compressive non-propagating
 mode.  The fact that the compressive mode has a higher power spectral
 density downstream of the high-$\beta$ than downstream of the
 low-$\beta$ shock might indicate the existence of mirror modes.
 We therefore perform a superposed epoch analysis for 9 cases for
 which the mirror instability criterion is particularly well
 satisfied, 7\,min downstream of the high-$\beta$ bow shock. The
 resulting spectrum is shown in Fig.~\ref{mmir}. In
 this spectrum we find a compressive mode slightly dominating at some
 frequencies well below the proton cyclotron frequency.

 
 Contrary to the correlation of the intensity of the left-hand
 polarized ion cyclotron waves to the proton temperature anisotropy
 for quasi-perpendicular low-$\beta$ cases, our investigation does not
 reveal a clear correlation of the intensity of the compressive mode
 to any parameter for the quasi-perpendicular high-$\beta$ cases.  One
 reason for this could be that the highly compressive mirror mode,
 which is expected to exist under the observed high-$\beta$ conditions
 is a purely growing mode with frequencies being pure Doppler-shifted
 frequencies.  Consequently, the waves do not appear in a fixed
 frequency interval and can be smeared out in the superposition.
 Therefore we have looked into the individual spectra of intervals 5,
 7, and 9\,min downstream of the quasi-perpendicular high-$\beta$ bow
 shock when the mirror criterion is fulfilled. This is the case in 34
 of the quasi-perpendicular high-$\beta$ events (72 \%). Only in 4
 cases (12 \% of the quasi-perpendicular high-$\beta$ events where the
 mirror criterion is fulfilled) mirror waves can clearly be identified
 in the magnetosheath and are visible in several consecutive spectra.
 These events and the means of identification of the mirror modes are
 described in {\it Czaykowska et al.} [1998]. The 4 events have in
 common that the angle $\theta_{Bn}$ is larger than $80^\circ$.  But
 there are also almost perpendicular high-$\beta$ events with large
 values of the left-hand side of Eq.~(\ref{mirin}) where no indication for
 mirror waves is visible.  Thus a more complex dependence on different
 parameters seems to determine the growth of the mirror wave. For 14
 of the high-$\beta$ events (41 \%) where the mirror instability
 criterion is fulfilled, the compressive component is at least
 slightly dominating at several frequencies. Several of these events
 have an angle $\theta_{Bn} < 80^\circ$.  In addition, 2
 high-$\beta$ events show ion cyclotron waves in the consecutive
 spectra taken 5, 7, and 9\,min downstream.  It is well known [{\it
   Price et al.}, 1986; {\it Gary et al.}, 1993] that the growths of
 the ion cyclotron and mirror waves are competing processes. In the
 magnetosheath plasma the crucial parameters for this competition are
 the $\alpha$-particle concentration and the plasma-$\beta$.  However,
 in our data set we have found many events in the high-$\beta$ regime
 where none of the two wave modes can be identified although the
 proton temperature anisotropy is high.  This seems to indicate that
 the dominance of the growing mode does not persist long enough to be
 visible in one spectrum. The energy of the growing mode might be
 transferred to other modes by nonlinear effects.

Statistical studies of measurements in the magnetosheath suggest that
a relation of the form 
\begin{equation}\label{anisobeta}
\frac{T_{p\perp}}{T_{p\|}} -1 = \frac{S}{(\beta_{p\|})^{\alpha}}
\end{equation}
exists between the proton temperature anisotropy and the ratio 
$\beta_{p\|}$ of field-aligned proton pressure and magnetic pressure.
Analyzing AMPTE/CCE data, {\it Anderson et al.} [1994] 
determined $S=0.85$ and $\alpha=0.48$ 
and {\it Fuselier et al.} [1994] determined $S=0.83$ and $\alpha=0.58$. 
Using AMPTE/IRM data, 
{\it Phan et al.} [1994] obtained $S=0.58$ and $\alpha=0.53$. 
We performed a similar analysis on our data set of quasi-perpendicular bow
shock crossings. For this analysis we computed 2-min averages of all 
measurements of $T_{p\perp}/T_{p\|}$ and $\beta_{p\|}$ 
taken between the keytime and 8\,min downstream of the keytime. 
We find a reasonable fit to Eq.~(\ref{anisobeta}) with 
$S=0.43\pm 0.03$ and $\alpha=0.58\pm 0.05$, which is not too different from the
result of {\it Phan et al.} [1994].

A relation of the form of Eq.~(\ref{anisobeta}) is regarded as the
consequence of the combined action of ion cyclotron and mirror waves,
which grow due to the temperature anisotropy and reduce this
anisotropy by means of pitch angle scattering. The growth of the waves
depends on $T_{p\perp}/T_{p\|}$ and $\beta_{p\|}$ and it is expected
that the anisotropy is reduced until the growth rate, $\gamma_m$, of
the most unstable wave falls below some threshold.  In fact, {\it
  Anderson et al.} [1994] showed that Eq.~(\ref{anisobeta}) with their
values of $S$ and $\alpha$ corresponds approximately to the threshold
$\gamma_m/2\pi f_{cp}=0.01$. Hence, the validity of
Eq.~(\ref{anisobeta}) indicates that the magnetosheath plasma reaches
a state near marginal stability of the waves driven by the temperature
anisotropy. As noted above, we found that data obtained less than
8\,min downstream of quasi-perpendicular bow shocks satisfy a relation
of the form Eq.~(\ref{anisobeta}) with values of $S$ and $\alpha$ that
are not too different from those determined by {\it Phan et al.}
[1994] for the entire magnetosheath.  This indicates that the state
near marginal stability is already reached close to the shock.

Figure~\ref{fluclh} shows that the largest amplitudes of the ion cyclotron 
waves at quasi-perpendicular low $\beta$ shocks are observed on average about 
5\,min downstream of the keytime (see also Fig.~\ref{corrver}). 
The bow shock moves relative to the spacecraft at speeds of 10--100\,km/s. 
Taking a typical speed of 30\,km/s, we can translate 5\,min to a downstream 
distance of 9000\,km. According to Fig.~\ref{ppsp1}, 
the plasma velocity, $v_{pn}$, normal to quasi-perpendicular shock is 120\,km/s 
on average. Thus the plasma needs about 75\,s to flow 9000\,km downstream. 
Since the ion cyclotron waves are convected with the plasma while they are
growing, these 75\,s can serve as a rough estimate for the time $\tau$ that 
the waves need to reach their maximum amplitudes and saturate. In terms of
gyro-periods, we have $\tau\sim 75\,{\rm s}\approx 50/f_{cp}$. 
Moreover, we find that on the same time scale $\tau$ the temperature
anisotropy is reduced from about 2.5 immediately downstream of the low $\beta$ 
shock to about 2.1 (Fig.~\ref{pplh}) and that 
the ion cyclotron waves typically reach amplitudes of $\delta B/B\sim 0.07$.

These results can be compared with two-dimensional hybrid simulations of
{\it McKean et al.} [1994]. These authors examined a plasma with 
$\beta_{p\|}=1$ and $T_{p\perp}/T_{p\|}=3$. 
Under these conditions the ion cyclotron mode is found
to be the dominant mode and the waves saturate after 
$\tau\approx5/f_{cp}$ and reach amplitudes of $\delta B/B\sim 0.15$. 
The proton temperature anisotropy is reduced on the same time scale $\tau$
from 3 to about 1.8. 
For our data set of low $\beta$ bow shocks 
$T_{p\perp}/T_{p\|}$ is on average 2.5 and $\beta_{p\|}$ is on average 0.5
immediately downstream of the shock. Since these values are considerably lower
than the initial values used by {\it McKean et al.} [1994], 
the plasma simulated by {\it McKean et al.} [1994] is initially much farther 
away from the state of marginal stability. 
Thus it is not surprising that the waves grow faster, reach larger 
amplitudes and therefore lead to a stronger reduction of the anisotropy by
means of pitch angle scattering. 

{\it McKean et al.} [1994] also examined a plasma with $\beta_{p\|}=4$ and 
$T_{p\perp}/T_{p\|}=3$. Under these conditions the ion cyclotron mode 
dominates for low $\alpha$-particle concentration,
whereas the mirror mode dominates for high $\alpha$-particle concentration.
The waves saturate after $\tau\approx 10/f_{cp}$ and reach 
amplitudes of $\delta B/B\sim 0.2$. 
The proton temperature anisotropy is reduced on the same time scale $\tau$
from 3 to about 1.5. 
This can be compared with data obtained at the quasi-perpendicular high $\beta$
shock. 
For our data set of high-$\beta$ bow shocks 
$T_{p\perp}/T_{p\|}$ is on average 1.8 and $\beta_{p\|}$ is on average 5
immediately downstream of the shock. 
Again, the plasma simulated by {\it McKean et al.} [1994] is initially 
much farther away from the state of marginal stability. 
Fig.~\ref{pplh} shows that a reduction of the anisotropy to 1.3 is observed 
30\,s downstream of the keytime. 
This can again be translated to a downstream distance and used to estimate the 
time span that passes while the plasma travels this distance. This estimate
gives $8\,{\rm s}\approx 2.5/f_{cp}$. 
Finally, it should be noted that the mirror waves analyzed by 
{\it Czaykowska et al.} [1998] 
have amplitudes of $\delta B/B\sim 0.2$, which is comparable to those found in 
the simulations of {\it McKean et al.} [1994].
 
\section{Conclusions}
We have analyzed the plasma and magnetic field parameters as well as
low frequency magnetic fluctuations at 132 dayside AMPTE/IRM bow shock
crossings. The average distance of the subsolar point, which results
from the coordinates of the investigated bow shock crossings, is
considerably smaller than in other studies, even when normalized to
the average solar wind dynamical pressure. A reason for this
discrepancy might be a variation of the polytropic index with the
solar cycle since our observations are performed during typical solar
minimum conditions. The position of the Earth's bow shock still seems
to be a matter of discussion.

A superposed epoch analysis has been carried out by averaging particle
and magnetic field data as well as low frequency magnetic spectra
upstream and downstream of the bow shock.
We have performed this analysis by dividing the events into
different categories, i.e., quasi-perpendicular and quasi-parallel
events as well as quasi-perpendicular low-$\beta$ and high-$\beta$
events.

The particle and magnetic field data show that upstream of the
quasi-parallel bow shock, in the foreshock region, the plasma is
already heated compared to the undisturbed solar wind. Moreover, there
are more energetic protons in the foreshock region, and the magnetic
field is highly variable.  Downstream of the quasi-perpendicular bow
shock, a proton temperature anisotropy is found, which is higher on
average downstream of the quasi-perpendicular low-$\beta$ than
downstream of the quasi-perpendicular high-$\beta$ bow shock.

Concerning the low frequency magnetic fluctuations we find that
upstream of the quasi-perpendicular bow shock the solar wind spectrum
is undisturbed with transverse Alfv\'en waves surpassing the
compressive spectral component. Upstream of the quasi-parallel bow
shock largely enhanced wave activity is detected in the turbulent
foreshock region. These upstream waves are convected downstream, 
experiencing an enhancement at the bow shock itself.
Downstream of the quasi-perpendicular bow shock the observed proton
temperature anisotropy leads to the generation of left-hand polarized
ion cyclotron waves under low-$\beta$ conditions and in some cases to
the generation of mirror waves under high-$\beta$ conditions.  A clear
correlation has been observed between the intensity of the left-hand
polarized component of the magnetic power spectrum relative to the
right-hand polarized component and the proton temperature anisotropy.
On the other hand, we could not find a simple correlation between the
intensity of the compressive component and any single plasma or
magnetic field parameter. In cases where mirror waves are obviously
observable mostly three conditions are fulfilled: the plasma-$\beta$
is high, the mirror instability criterion is satisfied and the angle
$\theta_{Bn}$ is large, i.e., $\theta_{Bn} \gtrsim 80^\circ$. But
there are also cases where all these conditions are well satisfied but
no mirror waves are visible in the frequency interval under
consideration.

\acknowledgments G. Paschmann and H. L\"uhr were the principal
investigators of the AMPTE/IRM plasma and magnetic field experiments,
respectively. We appreciate valuable discussions with N. Sckopke.

\end{document}